\documentclass[10pt,letterpaper,bibnotes,notitlepage,final,balancelastpage,superscriptaddress,twocolumn,showpacs,prl]{revtex4}
\usepackage{amssymb}
\usepackage{amsmath}
\usepackage{amsfonts}
\usepackage{graphicx}
\usepackage{color}

\begin{document}

\title{Efficient quantum repeater based on deterministic Rydberg gates}
\author{Bo Zhao}
\affiliation{Institute for Theoretical Physics, University of Innsbruck, A-6020
Innsbruck, Austria}
\affiliation{Institute for Quantum Optics and Quantum Information of the Austrian Academy
of Sciences, A-6020 Innsbruck, Austria}
\author{Markus M\"{u}ller}
\affiliation{Institute for Theoretical Physics, University of Innsbruck, A-6020
Innsbruck, Austria}
\affiliation{Institute for Quantum Optics and Quantum Information of the Austrian Academy
of Sciences, A-6020 Innsbruck, Austria}
\author{Klemens Hammerer}
\affiliation{Institute for Theoretical Physics, University of Innsbruck, A-6020
Innsbruck, Austria}
\affiliation{Institute for Quantum Optics and Quantum Information of the Austrian Academy
of Sciences, A-6020 Innsbruck, Austria}
\author{Peter Zoller}
\affiliation{Institute for Theoretical Physics, University of Innsbruck, A-6020
Innsbruck, Austria}
\affiliation{Institute for Quantum Optics and Quantum Information of the Austrian Academy
of Sciences, A-6020 Innsbruck, Austria}

\begin{abstract}
We propose an efficient quantum repeater architecture with mesoscopic atomic
ensembles, where the Rydberg blockade is employed for deterministic local
entanglement generation, entanglement swapping and entanglement
purification. Compared with conventional atomic-ensemble-based quantum
repeater, the entanglement distribution rate is improved by up to two orders
of magnitude with the help of the deterministic Rydberg gate. This new
quantum repeater scheme is robust and fast, and thus opens up a new way for
practical long-distance quantum communication.
\end{abstract}

\maketitle

Quantum information can be transmitted \textit{directly} over distances
above some hundred kilometers only at unpractically low rates due to loss
and decoherence. In order to remedy this limitation the concept of a quantum
repeater has been introduced \cite{Briegel98}, where quantum entanglement is
distributed over small distances, stored in quantum memories, purified, and
swapped in a nested architecture \cite{Van}. A quantum repeater can in
principle be implemented with atomic ensembles and linear optics only \cite%
{DLCZ}. However, despite significant progress during the last years on both
the theoretical \cite{Bo07,Sangouard08,Guo09} and experimental side \cite%
{Chou05,Zhao09}, -- see \cite{Sangouard09,Hammerer09} for recent reviews --
the entanglement distribution rate achievable in such an architecture is
still much too inefficient to be of practical interest, even under ideal
conditions. This is predominantly due to the fact that linear optical
methods only allow for a probabilistic entanglement manipulation, posing
severe limitations on the overall success probability, and therefore on the
rate of entanglement distribution.

In this letter we introduce a deterministic quantum repeater protocol using
quantum gates for entanglement swapping and purification. The quantum gates
rely on the Rydberg blockade effect in mesoscopic atomic ensembles \cite%
{Jaksch01}, and the remarkable recent advances in exploiting this effect for
quantum information processing \cite%
{Saffman07,Markus09,SaffmanNP09,Saffman10,Pfau09}. Deterministic operation
provides an enhancement of two orders of magnitude in the rate of
entanglement distribution as compared with the best quantum repeater based
on linear optics \cite{Sangouard08}. For realistic local errors around $%
10^{-3}-10^{-2}$ the new quantum repeater architecture yields a rate of
about 10 ebits per sec. We thus show that deterministic quantum repeaters
based on Rydberg gates open up a new avenue for high-rate, long-distance
quantum communication.

\begin{figure}[tbp]
\begin{center}
\includegraphics[
width=7cm]{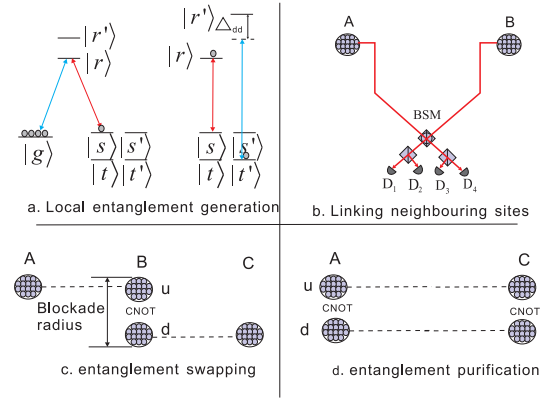}
\end{center}
\caption{A schematic view of the quantum repeater protocol with mesoscopic
atomic ensembles based on Rydberg gate, including local entanglement
generation, entanglement linking, entanglement swapping and entanglement
purification.}
\end{figure}

Deterministic quantum gates based on trapped ions were recently explored in
\cite{Simon09} in the context of quantum repeaters. However, this requires
strong coupling between single ions and high finesse cavities, which is
still challenging for current technology. The use of Rydberg gates in
quantum repeaters was first proposed in \cite{Molmer09}. In this protocol
the distribution of entanglement at the fundamental repeater level is still
probabilistic as it is based on absorption of photons which are lost in the
channel in most cases. The gate operation therefore has to rely on
post-selection, just as the conventional protocols \cite{Sangouard09} based
on linear-optics. In contrast, the quantum repeater architecture introduced
here is deterministic, does not require strong coupling between atoms and
light, and is robust against path length fluctuations.

In our protocol, mesoscopic atomic ensembles of the size of a few
micrometers are exploited as a quantum memory. If the atoms in such
ensembles are laser-excited to high-lying Rydberg states, strong and
long-range van-der-Waals or dipole-dipole interactions give rise to the
Rydberg blockade, which prevents the excitation of more than one Rydberg
atom within a volume, which is smaller than the blockade radius \cite%
{Jaksch01,Pfau09}. Based on the large nonlinearity associated with the
blockade effect deterministic entangling quantum gates can be performed
between collective excited states in one or different atomic ensembles by
applying a series of collective and single atom laser pulses \cite{Saffman07}%
. Our protocol starts by local and deterministic entanglement generation in
one atomic ensemble with the help of a collective Rydberg gate. The
entanglement is then linked between neighboring sites by linear optical
methods, where two photon interference is explored. Further entanglement
swapping and entanglement purification are implemented based on Rydberg
gates between two nearest memory atomic ensembles at one site. The protocol
presented here is improved in three respects compared with conventional
schemes: (i) local entanglement manipulation is performed deterministically,
(ii) the number of times required to convert atomic states into photons is
reduced to a minimum, (iii) the detection step in entanglement swapping and
entanglement purification can be performed with the help of field
ionization, thereby significantly increasing the detection efficiency.

We envision a setup with mesoscopic cold atomic ensembles with a diameter of
several microns. The relevant energy levels are shown Fig.~1a and comprise
an electronic ground state manifold with five sublevels $|g\rangle
,|s\rangle ,|s^{\prime }\rangle ,|t\rangle ,|t^{\prime }\rangle $, and two
Rydberg states which we denote by $|r\rangle $ and $|r^{\prime }\rangle $.
Initially all the atoms are prepared in the ground state $|g\rangle $. We
assume these sublevels can be addressed individually, and that atoms in the
two Rydberg states experience strong interactions.

In our scheme, we first generate a qubit-type entanglement in one atomic
ensemble, which can be done as follows (see Fig.~1a). i) A collective $\pi $
pulse (Rabi frequency $\Omega _{N}\propto \sqrt{N}$) and a single-atom $\pi $
pulse are applied sequentially to create one collective excitation,
transferring $|\mathbf{0}\rangle \rightarrow $ $|\mathbf{r}\rangle
\rightarrow |\mathbf{s}\rangle ,$ where $|\mathbf{0}\rangle =|g,...,g\rangle
$, and $|\mathbf{x}\rangle =\frac{1}{\sqrt{N}}\sum_{i=1}^N|g,\ldots, g,
x_{i}, g, \ldots, g \rangle $, where $x=r,r^{\prime },s,t,s^{\prime }$ or $%
t^{\prime }$ represents the collective state. In the intermediate step the
Rydberg blockade prevents excitation of more than one atom. ii) We create a
second collective excitation $|\mathbf{t}^{\prime }\rangle $ with the same
method. iii) A single-atom $\pi /2$ pulse transfers $|\mathbf{s}\rangle $ to
$(|\mathbf{s}\rangle +|\mathbf{r}\rangle )/\sqrt{2}.$ iv) A single-atom $\pi
$ pulse excites $|\mathbf{t}^{\prime }\rangle $ to $|\mathbf{r}^{\prime
}\rangle $. Due to dipole blockade, we obtain $(|\mathbf{s}\rangle |\mathbf{r%
}^{\prime }\rangle +|\mathbf{r}\rangle |\mathbf{t}^{\prime }\rangle )/\sqrt{2%
}.$ v) Finally, we apply two single-atom $\pi $ pulses to bring $|\mathbf{r}%
^{\prime }\rangle $ to $|\mathbf{s}^{\prime }\rangle $, and $|\mathbf{r}%
\rangle $ to $|\mathbf{t}\rangle $, and obtain the desired Bell state $(|%
\mathbf{s}\rangle |\mathbf{s}^{\prime }\rangle +|\mathbf{t}\rangle |\mathbf{t%
}^{\prime }\rangle )/\sqrt{2}.$

In a second step, all pairs of nearest communication sites are linked using
methods from linear optics \cite{Bo07,Sangouard08}: After the generation of
local entanglement at sites, say, A and B, read light pulses are applied to
convert the collective excitations in $|\mathbf{s}^{\prime }\rangle $ and $|%
\mathbf{t}^{\prime }\rangle $ into photons with different polarization,
e.g., $|H\rangle $ and $|V\rangle $ respectively, such that the whole system
is described by $(|\mathbf{s}_{A}\rangle |H_{A}\rangle +|\mathbf{t}%
_{A}\rangle |V_{A}\rangle )(|\mathbf{s}_{B}\rangle |H_{B}\rangle +|\mathbf{t}%
_{B}\rangle |V_{B}\rangle )/2$. The two photons from both sites are directed
to the middle point, and detected in a Bell state analyzer composed of a
polarizing beam splitter and single photon detectors \cite{Bo07}, where two
of the Bell states, e.g., $(|H_{A}\rangle |H_{B}\rangle \pm |V_{A}\rangle
|V_{B}\rangle )/\sqrt{2}$ are identified (see Fig. 1b). Once a two photon
coincidence count between the single photon detectors, e.g., D$_{1}$ and D$%
_{4}$, is registered, entanglement is generated between two memory qubits at
neighboring sites, described by $|\phi \rangle _{A,B}=(|\mathbf{s}%
_{A}\rangle |\mathbf{s}_{B}\rangle +|\mathbf{t}_{A}\rangle |\mathbf{t}%
_{B}\rangle )/\sqrt{2}$. This process is heralded, with a success
probability of $p=\frac{1}{2}\eta _{r}^{2}\eta _{pd}^{2}\eta _{att}^{2},$%
where $\eta _{r}$ is the retrieval efficiency, $\eta _{pd}$ is the photon
detection efficiency, and $\eta _{att}=e^{-L_{0}/(2L_{att})}$ denotes the
loss in the photonic channel with $L_{att}$ the attenuation length. If no
coincidence is registered, the local entanglement generation and linking
steps are repeated until success.

Finally, after neighboring communication sites are linked, we can connect
them by entanglement swapping. Suppose we have generated entanglement $|\phi
\rangle _{AB_{u}}$ and $|\phi \rangle _{B_{d}C}$ between atomic ensembles A
and B$_{u}$, B$_{d}$ and C, as shown in Fig. 1c. The two atomic ensembles at
site B are placed close to each other within the blockade radius, so that we
can perform a two-qubit gate between them. To implement entanglement
swapping, we first apply a CNOT gate between the memory qubits stored in
atomic ensembles B$_{u}$ and B$_{d}$, which can be done by a series of
single atom $\pi $ pulses \cite{Saffman05}: i) a $\pi $ pulse excites $|%
\mathbf{s}_{B_{u}}\rangle $ to $|\mathbf{r}_{B_{u}}\rangle ,$ ii) a $\pi $
pulse brings $|\mathbf{s}_{B_{d}}\rangle $ to $|\mathbf{r}_{B_{d}}\rangle $,
iii) a $\pi $ pulse transfers $|\mathbf{r}_{B_{d}}\rangle $ and $|\mathbf{t}%
_{B_{d}}\rangle $, iv) a $\pi $ pulse transfers $|\mathbf{r}_{B_{d}}\rangle $
to $|\mathbf{s}_{B_{d}}\rangle $, and v) a final $\pi $ pulse returns $|%
\mathbf{r}_{B_{u}}\rangle $ to $|\mathbf{s}_{B_{u}}\rangle .$ The
corresponding truth table is shown in Table 1.

\begin{table}[h]
\caption{Truth table of the CNOT gate operation between two ensembles
located at the same communication site, required for entanglement swapping.
The steps involving the Rydberg blockade mechanism are denoted by $%
\Rightarrow $.}
\begin{center}
$%
\begin{tabular}{|l|}
\hline
$\mathbf{s}_{{\small B}_{u}}\mathbf{s}_{{\small B}_{d}}{\small \rightarrow }%
\mathbf{r}_{{\small B}_{u}}\mathbf{s}_{{\small B}_{d}}{\small \Rightarrow }%
\mathbf{r}_{{\small B}_{u}}\mathbf{s}_{{\small B}_{d}}{\small \rightarrow }%
\mathbf{r}_{{\small B}_{u}}\mathbf{s}_{{\small B}_{d}}{\small \Rightarrow }%
\mathbf{r}_{{\small B}_{u}}\mathbf{s}_{{\small B}_{d}}{\small \rightarrow }%
\mathbf{s}_{{\small B}_{u}}\mathbf{s}_{{\small B}_{d}}$ \\ \hline
$\mathbf{s}_{{\small B}_{u}}\mathbf{t}_{{\small B}_{d}}{\small \rightarrow }%
\mathbf{r}_{{\small B}_{u}}\mathbf{t}_{{\small B}_{d}}{\small \rightarrow }%
\mathbf{r}_{{\small B}_{u}}\mathbf{t}_{{\small B}_{d}}{\small \Rightarrow }%
\mathbf{r}_{{\small B}_{u}}\mathbf{t}_{{\small B}_{d}}{\small \rightarrow }%
\mathbf{r}_{{\small B}_{u}}\mathbf{t}_{{\small B}_{d}}{\small \rightarrow }%
\mathbf{s}_{{\small B}_{u}}\mathbf{t}_{{\small B}_{d}}$ \\ \hline
$\mathbf{t}_{{\small B}_{u}}\mathbf{s}_{{\small B}_{d}}{\small \rightarrow }%
\mathbf{t}_{{\small B}_{u}}\mathbf{s}_{{\small B}_{d}}{\small \rightarrow }%
\mathbf{t}_{{\small B}_{u}}\mathbf{r}_{{\small B}_{d}}{\small \rightarrow }%
\mathbf{t}_{{\small B}_{u}}\mathbf{t}_{{\small B}_{d}}{\small \rightarrow }%
\mathbf{t}_{{\small B}_{u}}\mathbf{t}_{{\small B}_{d}}{\small \rightarrow }%
\mathbf{t}_{{\small B}_{u}}\mathbf{t}_{{\small B}_{d}}$ \\ \hline
$\mathbf{t}_{{\small B}_{u}}\mathbf{t}_{{\small B}_{d}}{\small \rightarrow }%
\mathbf{t}_{{\small B}_{u}}\mathbf{t}_{{\small B}_{d}}{\small \rightarrow }%
\mathbf{t}_{{\small B}_{u}}\mathbf{t}_{{\small B}_{d}}{\small \rightarrow }%
\mathbf{t}_{{\small B}_{u}}\mathbf{r}_{{\small B}_{d}}{\small \rightarrow }%
\mathbf{t}_{{\small B}_{u}}\mathbf{s}_{{\small B}_{d}}{\small \rightarrow }%
\mathbf{t}_{{\small B}_{u}}\mathbf{s}_{{\small B}_{d}}$ \\ \hline
\end{tabular}%
\ $%
\end{center}
\end{table}

After applying the CNOT gate we measure the memory qubits in the ensembles B$%
_{u}$ and B$_{d}$ in four states $|\mathbf{+}_{B_{u}}\rangle |\mathbf{s}%
_{B_{d}}\rangle ,|\mathbf{-}_{B_{u}}\rangle |\mathbf{s}_{B_{d}}\rangle ,|%
\mathbf{+}_{B_{u}}\rangle |\mathbf{t}_{B_{d}}\rangle $ and $|\mathbf{-}%
_{B_{u}}\rangle |\mathbf{t}_{B_{d}}\rangle ,$ where $|\mathbf{\pm }%
_{B_{u}}\rangle =(|\mathbf{s}_{B_{u}}\rangle \pm |\mathbf{t}_{B_{u}}\rangle
)/\sqrt{2},$ in order to project the memory qubits at sites A and C into the
desired entangled state. In contrast to conventional schemes where the
collective excitations are converted into photons for the detection, we
suggest to measure the quantum state by transferring the excitation to a
Rydberg state, field-ionizing the atom and detecting the ions. Detection of
single Rydberg atoms has been demonstrated in photon counting experiment
with near-unity detection efficiencies $\eta_{d}$ \cite{Haroche07}. After
the detection, the states are projected into $(|\mathbf{s}_{A}\rangle |%
\mathbf{s}_{C}\rangle +|\mathbf{t}_{A}\rangle |\mathbf{t}_{C}\rangle )/\sqrt{%
2}$, up to a local unitary transformation.

The communication distance can be extended further by entanglement swapping.
Since entanglement swapping is deterministic, the entanglement distribution
rate is similar to the one of a quantum repeater based on trapped ions \cite%
{Simon09}. For $L=2^{n}L_{0}$, the total time needed can be approximated by
\begin{equation*}
T_{tot}\approx \prod\limits_{i=0}^{n}\alpha _{i}\frac{T_{cc}}{p}\approx
\frac{\prod\limits_{i=1}^{n}\alpha _{i}}{2^{n-1}}\frac{3L}{\eta _{r}^{2}\eta
_{pd}^{2}e^{-L/(2^{n}L_{att})}}
\end{equation*}%
where $T_{cc}=\frac{L_{0}}{c}$ the classical communication time with $c$ the
light speed, $p=\frac{1}{2}\eta _{r}^{2}\eta _{pd}^{2}\eta _{t}^{2}$ and $%
\alpha _{0}/p$ are are the average of times one has to repeat before
entanglement is successfully linked over the entire distance, with a
numerical result of $\alpha _{0}\approx 3$ for $p\ll 1$ \cite{note}. The
coefficients $\alpha _{i\neq 0}>1$ denote the average number of attempts
needed to implement entanglement swapping due to non-unity detection
efficiency.

\begin{figure}[tbp]
\begin{center}
\includegraphics[width=\columnwidth]{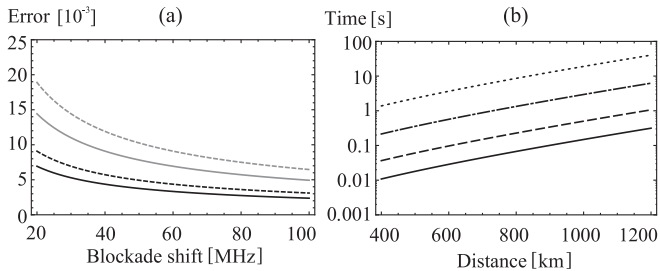}
\end{center}
\caption{(a) Average errors in local entanglement generation (gray) and
entanglement swapping (black) versus dipole-dipole shift. The dashed and
solid curves are for $\protect\tau =200$ and 300 $\protect\mu $s
respectively. (b) The performance of the quantum repeater. The solid curve
represents the result without entanglement purification, which requires
local errors on the order of $10^{-3}$. The dashed and dotted dashed line
are the results for $F_{loc}=F_{cnot}=0.99$ and $0.98$, where active
entanglement purification is implemented twice and four times respectively,
when the fidelity is no larger than $0.9$. The final fidelity is higher than
$0.94$ and the probability to get the entangled state is larger than $0.95$
in both cases. The dotted line is the result of the best-known protocol with
atomic ensembles and linear optics proposed in Ref. \protect\cite%
{Sangouard08}. }
\end{figure}

Let us now take into account local manipulation errors, which have been
neglected in the discussion so far. The intrinsic errors in local
manipulation are mainly induced by decay of the atoms when they are excited
to Rydberg states, the imperfect Rydberg blockade induced by finite
dipole-dipole shifts \cite{Saffman05}, and the imprecision of the collective
pulses caused by an uncertainty of the atom number $N$. The decay of the
Rydberg states causes decoherence errors proportional to the Rydberg states
decay rate $\gamma $ and inversely proportional to the Rabi frequency of the
collective pulses $\Omega _{N}$ or single atom pulses $\Omega _{s}$. The
finite value of the Rydberg interaction energy shift $\Delta _{dd}$
(imperfect blockade) will cause several kinds of errors. The first one is
that two excitations may be generated in the atomic ensembles and thus
causes losses. Secondly, the adiabatic elimination of the doubly excited
Rydberg states will cause an ac stark shift on the one atom excitation
states, and thus cause dephasing errors. These errors in the local
entanglement generation step are of the order $\Omega_{N,s}^{2}/\Delta
_{dd}^{2}$ and can be estimated as $E_{loc}=1-F_{loc}=2\frac{\gamma \pi }{%
\Omega _{N}}+\frac{\gamma \pi }{\Omega _{s}}+4\frac{\Omega _{N}^{2}}{\Delta
_{dd}^{2}}+2\frac{\Omega _{s}^{2}}{\Delta _{dd}^{2}}$, with $F_{loc}$ the
fidelity of the locally achieved entanglement. The imprecision of the
collective $\pi $ pulses is on the order of $1/N$ $\ $for an uncertainty of
the atom numbers $\sqrt{N}$. For $N>100$, this error is less than $1\%$ and
can be safely neglected. In Fig. 2a we plot the optimized local errors $%
E_{loc}$ versus $\Delta _{dd}$ for $\tau =1/(2\pi \gamma )=200$ and $300$ $%
\mu $s (and $\Omega _{N}=\Omega _{s}$). One can see the local error is only
a few percent for a dipole shift of $\Delta _{dd}=20-100$ MHz.

Local imperfections are mainly decoherence, dephasing and loss errors. We
can thus neglect spin flip errors and describe the local entanglement by a
mixed entangled state $\rho =(1-(p_{1}+p_{0}))\rho _{2}+p_{1}\rho
_{1}+p_{0}\rho _{0}),$ where $\rho _{2}=F_{loc}|\phi ^{+}\rangle \langle
\phi ^{+}|+(1-F_{loc})|\phi ^{-}\rangle \langle \phi ^{-}|$ with $|\phi
^{\pm }\rangle =(|\mathbf{s}_{A}\rangle |\mathbf{s}_{B}\rangle \pm |\mathbf{t%
}_{A}\rangle |\mathbf{t}_{B}\rangle )/\sqrt{2}$ , $p_{1}\sim $ $%
(\Omega_{N}/\Delta _{dd})^{2}$ and $p_{0}\sim(\Omega _{s}/\Delta _{dd})^{2}$
are respectively the small probabilities to generate erroneously a single
excitation and vacuum contribution $\rho _{1}$ and $\rho _{0}$, which are
created due to double excitations (imperfect Rydberg blockade).

After linking the neighboring sites, we obtain a density matrix $\rho
^{0}=(1-O(p_{1})\rho _{2}^{0}+O(p_{1})\rho _{1})$ up to the first order of $%
O(p_{1})$, with a success probability of $p\approx \frac{1}{2}\eta
_{r}^{2}\eta _{pd}^{2}e^{-L_{0}/L_{att}}(1-O(p_{1})),$ where $\rho
_{2}^{0}=F_{0}|\phi ^{+}\rangle \langle \phi ^{+}|+(1-F_{0})|\phi
^{-}\rangle \langle \phi ^{-}|$ up to a local unitary transformation, with $%
F_{0}=F_{loc}^{2}+(1-F_{loc})^{2}.$ The errors in the photonic channel are
neglected since two photon interference is used. The errors in the
subsequent entanglement swapping step are similar to the ones of the local
entanglement generation, and can be estimated using the average error of the
CNOT gate $E_{cnot}=1-F_{cnot}=\frac{2\gamma \pi }{\Omega _{s}}+\frac{%
3\Omega _{s}^{2}}{2\Delta _{dd}^{2}}$. Figure 2a shows the optimized
swapping errors $E_{cnot}$ versus $\Delta _{dd}$ for $\tau =1/(2\pi \gamma
)=200$ and $300$ $\mu $s. The entanglement swapping errors are smaller than
for the generation of local entanglement since no collective pulses with
associated generation of collective excitations are required.

After $n$-step entanglement swapping, the mixed entangled state reads $\rho
^{n}=(1-O(p_{1})\rho _{2}^{n}+O(p_{1})\rho _{1})$ where $\rho
_{2}^{n}=F_{n}|\phi ^{+}\rangle \langle \phi ^{+}|+(1-F_{n})|\phi
^{-}\rangle \langle \phi ^{-}|.$ The fidelity can be approximated by $%
F_{n}=(F_{n-1}^{2}+(1-F_{n-1})^{2})F_{cnot}$, for $F_{cnot}$ close to one.
Note that the probability of obtaining the two excitations is independent of
$n$, thanks to the use of qubit-type entanglement and the detection of two
excitations in each step \cite{Bo07}. The success probability of each
entanglement swapping step is $p\approx (1-O(2p_{1}))\eta _{d}^{2}$, where
we have assumed for simplicity that the probability to obtain a double
Rydberg excitation during the entanglement swapping $(\Omega _{s}/\Delta
_{dd})^{2}$ is on the order of $O(p_{1})$.

The local errors will accumulate during entanglement connection \cite%
{Briegel98}, and thus entanglement purification has to be performed which
can be achieved by employing two CNOT gates \cite{Bennett96}. Assume we have
generated two pairs of mixed entangled states between A$_{u}$ and C$_{u}$,
and A$_{d}$ and C$_{d}$, described by $\rho ^{i}$. We first apply a $\pi /2$
Raman pulse coupling $|\mathbf{s}\rangle $ and $|\mathbf{t}\rangle $ to
change the two excitation component to $\rho _{2}^{i\prime }=F_{i}|\phi
^{+}\rangle \langle \phi ^{+}|+(1-F_{i})|\psi ^{+}\rangle \langle \psi ^{+}|$
with $|\psi ^{+}\rangle =(|\mathbf{s}_{A_{u,d}}\rangle |\mathbf{t}%
_{C_{u,d}}\rangle +|\mathbf{t}_{A_{u,d}}\rangle |\mathbf{s}_{C_{u,d}}\rangle
)/\sqrt{2}$. We then perform two local CNOT gates with A$_{u}$ and C$_{u}$
the control qubit and A$_{d}$ and C$_{d}$ the target qubits, where we have
assumed the two atomic ensembles at one site are located within the blockade
radius. After the CNOT gates, we measure the target qubits in ensembles A$%
_{d}$ and C$_{d}$ in the $|\mathbf{s}\rangle $ and $|\mathbf{t}\rangle $
basis. If both qubits are in the $|\mathbf{s}\rangle $ or the $|\mathbf{t}%
\rangle $ state, the memory qubits in A$_{u}$ and C$_{u}$ are kept,
otherwise the results are discarded. After entanglement purification, we
obtain a mixed state $\rho _{2}^{p}=F^{p}|\phi ^{+}\rangle \langle \phi
^{+}|+(1-F^{p})|\psi ^{+}\rangle \langle \psi ^{+}|$, where the leakage to
other states is neglected for large values of $F_{cnot}$, and the achieved
fidelity can be estimated as $F^{p}=\frac{F_{i}^{2}}{F_{i}^{2}+(1-F_{i})^{2}}%
F_{cnot}^{2}$. The success propagability of purification is $p\approx
(F_{i}^{2}+(1-F_{i})^{2})\eta _{d}^{2}.$ After entanglement purification,
the total density matrix can be described by $\rho
=(1-O(p_{1}/F_{i}^{2})\rho _{2}^{p}+O(p_{1}/F_{i}^{2})\rho _{1})$ where we
have assumed that the one excitation term only contributes a false signal.

The main result of our work is illustrated in Fig. 2b, where the performance
of the quantum repeater is plotted as a function of the communication
distance for $n=4,$ $\eta _{r}=\eta _{pd}=0.9,\eta _{d}=0.95,$\ $L_{att}=22$
km and $c=2\times 10^{5}$ km/s in fibers. For comparison, we also show the
performance of the best known atomic-ensemble-based repeater protocol
without purification \cite{Sangouard08}. It can be seen that the
entanglement distribution rate is enhanced by up to two orders of magnitude.
For $L=1000$ km, the total time needed is on the order of a few hundred
milliseconds.

The presented quantum repeater can be implemented using cold alkali atoms.
Individual addressing of different sublevels can be achieved by choosing
suitable laser polarization and applying a constant magnetic field. In our
protocol, we suggest to use the isotropic repulsive van der Waals
interactions by exciting the atoms to Rydberg to $s$-states with a principal
quantum number $n$ around $70$. In this case, the interaction energy between
two atoms at a distance of $r$ can be approximated by $V=-c_{1}\frac{n^{12}}{%
r^{6}}+$ $c_{1}^{\prime }\frac{n^{16}}{r^{8}}$ \cite{Weidemuller05}, with $%
c_{1}<0$ and $c_{1}^{\prime }>0$, where interactions proportional to $%
1/r^{10}$ are neglected. The interactions are repulsive for large and
attractive for small $r$, yielding a critical distance $r_{c}$ where the
repulsive shift is maximal. We use $r_{c}$ to estimate the minimum distance
required to assure repulsive interatomic interactions, and find for Rb, $%
c_{1}=-0.85$ and $c_{1}^{\prime }=0.8$, and $n=70,$ a critical distance $%
r_{c}=0.3$ $\mu $m, corresponding to a density of $1/(r_{c}^{3})=3.7\times
10^{13}/$cm$^{3}.$ For a fixed density, one is interested in maximizing the
number $N$ of atoms within the blockade radius, for high photon retrieval
efficiencies and a uncertainty in the atom number. As illustrated in Fig.
2a, an interaction energy shift $\Delta _{dd}>20$ MHz allows for local
errors of less than $2\%$. This yields a maximum Rydberg blockade radius $%
R_{b}\approx (-\frac{c_{1}n^{12}}{\Delta _{dd}})^{1/6}$; a more accurate
calculation using the interaction energy in \cite{Weidemuller05} gives $%
R_{b}<6$ $\mu $m for $n=70$. Thereby, a diameter of $2$ to $3$ $\mu $m is
sufficient for achieving high fidelity local operations (a density of $%
3\times 10^{13}/$cm$^{3}$ and a volume of ($2$ $\mu $m$)^{3}$ would allow
for about $N=240$ atoms per ensemble). With the help of a bad cavity, the
retrieval efficiency can be estimated as $\eta _{r}=\frac{C}{C+1}$, where $%
C=Nc_{r}^{2}\frac{24F}{2\pi k^{2}w_{0}^{2}}$ with $k$ the wave number of the
emitted photon, and $c_{r}$ the transition coefficient \cite{SimonJ07}. For
a finesse of $F=100$, cavity mode width $w_{0}=5$ $\mu $m, $c_{r}=\frac{1}{3}
$ and $k=2\pi /\mu m$, we can obtain a high retrieval efficiency of $0.91.$

Finally, to implement long distance quantum communication over 1000 km, the
coherence times of the quantum memory have to be on the order of a few
hundred milliseconds. This should be achievable for an atomic memory with
cold atoms, where a storage time of about one second for classical light has
been achieved \cite{Hau09}.

Financial support by the Austrian Science Foundation (FWF) through SFB
FOQUS, and the EU projects SCALA and NAMEQUAM is acknowledged.


\begin{thebibliography}{99}
\bibitem{Briegel98} H.-J. Briegel \textit{et al.}, Phys. Rev. Lett. \textbf{%
81}, 5932 (1998).

\bibitem{Van} L. Childress \textit{et al.}, Phys. Rev. Lett. \textbf{96},
070504 (2006); P. van Loock \textit{et al.}, \textit{ibid.} \textbf{96},
240501 (2006).

\bibitem{DLCZ} L.M. Duan \textit{et al.}, Nature \textbf{414}, 413 (2001).

\bibitem{Bo07} B. Zhao \textit{et al.}, Phys. Rev. Lett. \textbf{98}, 240502
(2007); Z.-B. Chen \textit{et al.}, Phys. Rev. A \textbf{76}, 022329 (2007);
L. Jiang \textit{et al.}, \textit{ibid.} \textbf{76}, 012301 (2007).

\bibitem{Sangouard08} N. Sangouard \textit{et al.}, Phys. Rev. A \textbf{77}%
, 062301 (2008).

\bibitem{Guo09} M. Gao \textit{et al.}, Phys. Rev. A \textbf{79}, 042301
(2009); Z.-Q. Yin \textit{et al.}, \textit{ibid.} \textbf{79}, 044302 (2009).

\bibitem{Chou05} D. N. Matsukevich and A. Kuzmich, Science \textbf{306}, 663
(2004); C.-W. Chou \textit{et al.}, Nature \textbf{438}, 828 (2005); C.-W.
Chou \textit{et al.}, Science \textbf{316}, 1316 (2007); Y.-A. Chen \textit{%
et al.}, Nat. Phys. \textbf{4}, 103 (2008); Z.-S. Yuan \textit{et al.},
Nature \textbf{454}, 1098 (2008).

\bibitem{Zhao09} B. Zhao \textit{et al.}, Nat. Phys. \textbf{5}, 95 (2009);
R. Zhao, \textit{et al.}, \textit{ibid.} \textbf{5}, 100 (2009).

\bibitem{Sangouard09} N. Sangouard \textit{et al.}, arXiv:0906.2699 (2009).

\bibitem{Hammerer09} K. Hammerer \textit{et al.}, arXiv:0807.3358 (2008).

\bibitem{Jaksch01} D. Jaksch \textit{et al.}, Phys. Rev. Lett. \textbf{85},
2208 (2000); M. D. Lukin \textit{et al.}, \textit{ibid.} \textbf{87}, 037901
(2001).

\bibitem{Saffman07} E. Brion \textit{et al.}, Phys. Rev. Lett. \textbf{99},
260501 (2007).

\bibitem{Markus09} M. M\"{u}ller \textit{et al.}, Phys. Rev. Lett. \textbf{%
102}, 170502 (2009).

\bibitem{SaffmanNP09} E. Urban \textit{et al.}, Nat. Phys. \textbf{5}, 110
(2009); A. Ga\"{e}tan \textit{et al.}, \textit{ibid.} \textbf{5}, 115 (2009).

\bibitem{Saffman10} T. Wilk \textit{et al.}, \textit{ibid.}, \textbf{104},
010502 (2010); L. Isenhower \textit{et al.}, Phys. Rev. Lett. \textbf{104},
010503 (2010).

\bibitem{Pfau09} C. Liebisch \textit{et al.}, Phys. Rev. Lett. \textbf{95},
253002 (2005); A. K. Mohapatra \textit{et al.}, \textit{ibid.} \textbf{98},
113003 (2007); H. Weimer \textit{et al.}, \textit{ibid.} \textbf{101},
250601 (2008); H. K\"{u}bler \textit{et al.}, arXiv:0908.0275v1 (2009).

\bibitem{Simon09} N. Sangouard \textit{et al.}, Phys. Rev. A \textbf{79},
042340 (2009).

\bibitem{Molmer09} L. H. Pedersen and K. M{\o }lmer, Phys. Rev. A \textbf{79}%
, 012320 (2009).

\bibitem{Haroche07} C. Guerlin \textit{et al.}, Nature \textbf{448}, 889
(2007).

\bibitem{note} We first randomly generate $m$ numbers according to a
Bernoulli distribution with a success probability of $p$. We then generate $%
m^{\prime } $ numbers according to a Bernoulli distribution, where $%
m^{\prime }=m-k$ with $k$ the number of "$1$" events in the last step. This
procedure is repeated until finally $m^{\prime }=0$. The average number of
times needed gives $\alpha _{0}/p$. The coefficients $\alpha _{i\neq 0}$ are
determined similarly.

\bibitem{Saffman05} M. Saffman and T. G. Walker, Phys. Rev. A \textbf{72},
022347 (2005).

\bibitem{Bennett96} C. H. Bennett \textit{et al.}, Phys. Rev. Lett \textbf{76%
}, 722 (1996); D. Deutsch \textit{et al.}, \textit{ibid.}, \textbf{77}, 2818
(1996).

\bibitem{Weidemuller05} K. Singer \textit{et al.}, J. Phys. B \textbf{38},
S295 (2005).

\bibitem{SimonJ07} J. Simon \textit{et al.}, Phys. Rev. Lett. \textbf{98},
183601 (2007).

\bibitem{Hau09} R. Zhang \textit{et al.}, Phys. Rev. Lett. \textbf{103},
233602 (2009); U. Schnorrberger \textit{et al.} \textit{ibid.}, \textbf{103}%
, 033003 (2009).
\end{thebibliography}
\end{document}